\documentclass[%
 aip,
 amsmath,amssymb,
 reprint,%
]{revtex4-1}

\usepackage{graphicx}
\usepackage{dcolumn}
\usepackage{bm}

\usepackage[utf8]{inputenc}
\usepackage[T1]{fontenc}
\usepackage{mathptmx}
\usepackage{etoolbox}

\usepackage[english]{babel}

\usepackage{siunitx}

\usepackage{csquotes}
\MakeOuterQuote{"}

\usepackage{bm}
\usepackage{braket}
\renewcommand{\v}[1]{\bm{\mathrm{#1}}}
\newcommand{\m}[1]{\bm{\mathsf{#1}}}

\makeatletter
\def\@email#1#2{%
 \endgroup
 \patchcmd{\titleblock@produce}
  {\frontmatter@RRAPformat}
  {\frontmatter@RRAPformat{\produce@RRAP{*#1\href{mailto:#2}{#2}}}\frontmatter@RRAPformat}
  {}{}
}%
\makeatother

\begin{document}

\preprint{AIP/123-QED}

\title[Creation and control of valley currents in graphene by few cycle light pulses]{Creation and control of valley currents in graphene by few cycle light pulses}

\author{Deepika Gill}
 \affiliation{Max-Born-Institut f\"ur Nichtlineare Optik und Kurzzeitspektroskopie, Max-Born-Strasse 2A, 12489 Berlin, Germany.}
 
\author{Sangeeta Sharma}
 \email{sharma@mbi.de}
 \affiliation{Max-Born-Institut f\"ur Nichtlineare Optik und Kurzzeitspektroskopie, Max-Born-Strasse 2A, 12489 Berlin, Germany.}
\affiliation{Institute for theoretical solid-state physics, Freie Universit\"at Berlin, Arnimallee 14, 14195 Berlin, Germany
}

\author{Sam Shallcross}
\affiliation{Max-Born-Institut f\"ur Nichtlineare Optik und Kurzzeitspektroskopie, Max-Born-Strasse 2A, 12489 Berlin, Germany.}

\date{\today}

\begin{abstract}
Well established for the visible spectrum gaps of the transition metal dichalcogenide family, valleytronics -- the control of valley charge and current by light -- is comparatively unexplored for the THz gaps that characterize graphene and topological insulators. Here we show that few cycle pulses of THz light can create and control a 100\% valley polarized current in graphene, with lightwave control over the current magnitude and direction. The latter is equal to an emergent pulse property of few cycle circularly polarized pulses,  the "global" carrier envelope phase. Our findings both highlight the richness of few cycle light pulses in control over quantum matter, and provide a route towards a "THz valleytronics" in meV gapped systems.
\end{abstract}

\maketitle

Novel quantum degrees of freedom, such as valley state and topological charge, provide rich new possibilities for controlling matter by light, and potentially form key elements of future classical and quantum computing technologies. The valley degree of freedom, that directly couples to circularly polarized light, has in particular attracted attention, most notably in the context of transition metal dichalcogenide (TMDC) family\cite{xiao_nonlinear_2015,mak_lightvalley_2018,
langer_lightwave_2018,vitale_valleytronics_2018,
li_room-temperature_2020} whose valley gaps of 1.6-2.3 eV fall within the visible spectrum. Terahertz light\cite{bera_review_2021,mashkovich_terahertz_2021,
khan_ultrafast_2020,chekhov_ultrafast_2021} -- typically understood as light in the 1-30 THz frequency range -- couples to several fundamental electronic and lattice energy scales of the solid state, allowing for example light enhancement of Curie temperature and control over magnetic order\cite{disa_polarizing_2020,disa_photo-induced_2023,curtis_dynamics_2023}. THz valleytronics, by contrast, remains relatively unexplored despite the existence of materials with excellent valleytronics potential that possess gaps in the THz window, for example graphene encapsulated in h-BN\cite{yankowitz_van_2019} and topological insulator (TI) surface states.

A direct analogy of the valley charge polarization physics of the TMDC family fails for THz gaps, as thermal fluctuations wash out light created valley contrast. However, valley current remains as a physical property that can be created and controlled by light in "THz valleytronics", with "hencomb" pulse designs\cite{sharma_giant_2023} generating 70-80\% valley polarization of current. Circularly polarized light is generally thought of as coupling only to the charge state of valleys, as the rotating polarization vector cannot distinguish directions in momentum space. However, recent work has shown that this picture breaks down in the few cycle limit\cite{sharma_direct_2024} for the TMDC family: at such short times the lightform becomes inherently vectorial and can thus simultaneously control valley charge and valley current.

Here we show that this short time light-matter symmetry breaking of valley physics holds also for THz gaps. Taking as a test case graphene with a 40~meV gap -- typical of those found upon encapsulation in h-BN\cite{yankowitz_van_2019} -- we demonstrate the generation of nearly 100\% pure valley current by few cycle pulses of THz light with complete direction control via the "global" carrier envelope phase of the pulse. While generation of valley current in graphene is typically discussed in the context of complex nano-device structures -- for example valley filters\cite{settnes_graphene_2016,
yesilyurt_perfect_2016,faria_valley_2020,
rycerz_valley_2007,yao_line-defectinduced_2015,khalifa_weyl_2021,
jo_quantum_2021} employ the emergent gauge physics of deformed graphene\cite{levy_strain-induced_2010,nigge_room_2019,
gupta_straintronics_2019,pacakova_mastering_2017,
hsu_nanoscale_2020,nigge_room_2019} -- our work provides a contrasting route to valleytronics in graphene via the short time limit of circularly polarized laser pulses.



\begin{figure*}[t!]
\begin{center}
\includegraphics[width=0.7\textwidth]{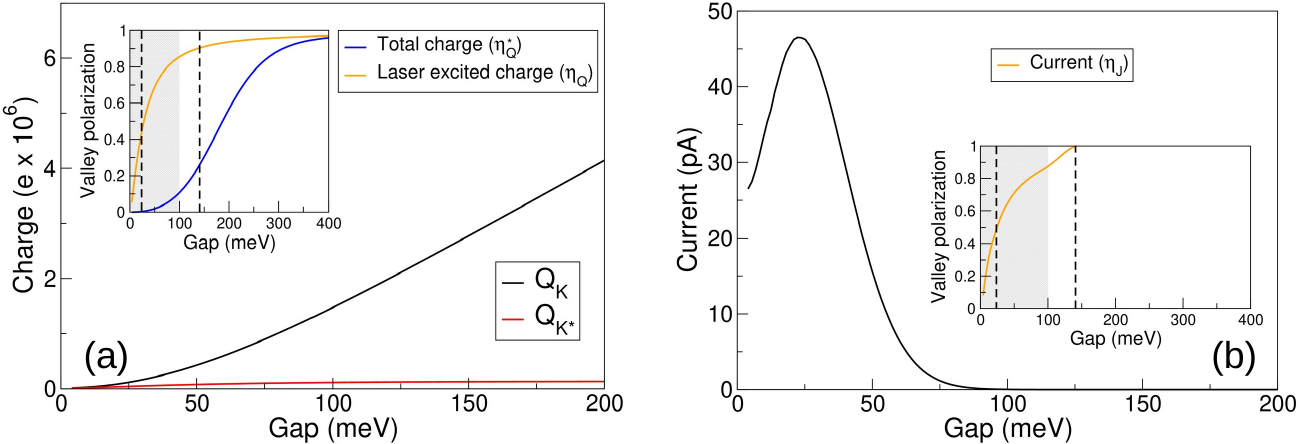}
\caption{\small {\it In plane symmetry breaking by circularly polarized THz light in gapped graphene.} (a) Charge excited at the conjugate K and K$^\ast$ valleys by circularly polarized light pulse tuned to the gap. Valley polarization by light fails as the gap vanishes, a fact confirmed by calculation of the valley polarization $\eta_Q = (Q_K - Q_{K^*})/(Q_K + Q_{K^*})$, see inset. Including thermally excited charge at $T=300$~K in the valley charge leads to a significantly earlier loss of valley polarization, presented as $\eta_Q^*$ in the inset. (b) The low gap region in which valley charge polarization fails is, however, endowed with an emergent current not found at large gaps. This current exhibits valley polarization corresponding to the underlying charge polarization, panel (b) inset.
}
\label{fig1}
\end{center}
\end{figure*}

We consider a circularly polarized light pulse applied to a graphene system in which a small gap $\Delta$ has been imposed at the Dirac point. Such gaps can readily be introduced by the particular environment in which the graphene layer is situation, for example by encapsulation in hexagonal Boron Nitride (h-BN)\cite{dean_boron_2010,
yankowitz_van_2019}, or epitaxial growth on the (0001) face of SiC\cite{zhou_substrate-induced_2007}. To this end we deploy a minimal model consisting of the nearest neighbour tight-binding approach with $\Delta$ introduced via a mass field, treating the dynamical evolution induced by an ultrafast laser pulse via the von Neumann equation incorporating a phenomenological decoherence of 20~fs\cite{heide_electronic_2021}. Further details of our model and its numerical solution are provided in Appendix A. 

\begin{figure}[t!]
\begin{center}
\includegraphics[width=0.45\textwidth]{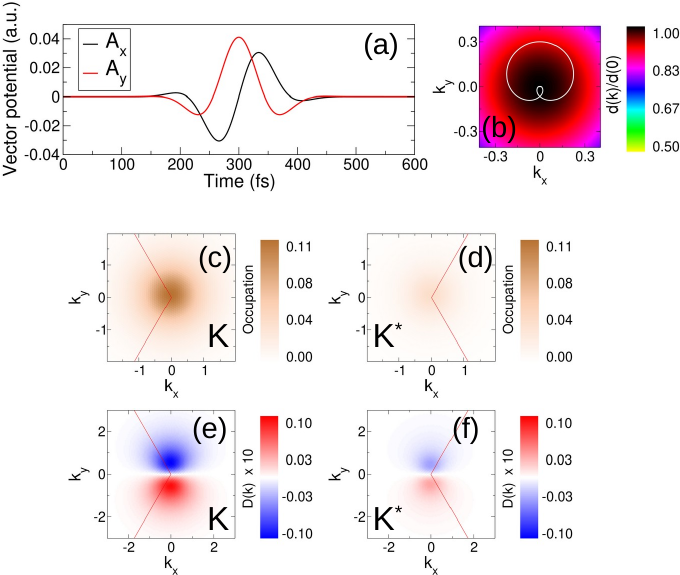}
\caption{\small {\it Emergence of light-matter symmetry breaking in the single cycle limit of circularly polarized light}. (a) The vector potential of a single cycle THz pulse, tuned to a gap of 30~meV (7.2~THz). The dynamical trajectory i.e. the evolution of crystal momentum induced by the lightform, describes a single loop in momentum space, panel (b). This generates symmetry breaking of the light-matter coupling that while difficult to discern in the momentum resolved charge density, panels (c,d), is clearly visible in the corresponding symmetry breaking density $D(\v k)$, Eq.~\ref{DK}, panels (e,f), that reveals the light induced excitation to be symmetry lowered from the underlying $C_3$ symmetry of the band manifold to the $C_2$ symmetry of the single loop lightform.
}
\label{fig2}
\end{center}
\end{figure}

Circularly polarized light fails as a route to valley polarization in the small gap limit. This is illustrated in Fig.~\ref{fig1}(a) in which the charge at the K valley and its conjugate K$^\ast$ partner is shown as a function of the graphene gap. The light pulse in each case has fixed duration and amplitude, 117~fs and 0.041~a.u. respectively, with the frequency tuned to the gap. While for $\Delta > 100$~meV clear valley polarization can be seen, with light inducing dramatically more charge at the K as compared to K$^\ast$ valley, this falls to zero as the gap enters the "THz window", a fact confirmed by a calculation of the valley polarization $\eta_Q = (Q_K - Q_{K^*})/(Q_K + Q_{K^*})$, see inset panel. Unexpectedly, however, this small gap region exhibits an increase of valley, Fig.~\ref{fig2}(b).  Calculation of the analogously defined valley current polarization, $\eta_J = (J_K - J_{K^*})/(J_K + J_{K^*})$, shows furthermore that this current is strongly valley polarized, in particular towards the weak current tail at gaps of 50-100~meV, see inset panel of Fig.~\ref{fig2}(b).

To explore the origins of this we consider in detail the case of a 30~meV gap corresponding to the maximal current response. The current carrying capacity of the charge excitations induced by a light pulse can be accessed by introducing a symmetry breaking density\cite{sharma_direct_2024}:

\begin{equation}
D(\v k) = |c_{\v k}|^2 - \frac{1}{3}
\sum_{i=1}^{3} |c_{M_i \v k}|^2.
\label{DK}
\end{equation}
The value of $D(\v k)$ represents the deviation of the conduction band occupation at crystal momenta $\v k$, $|c_{\v k}|^2$, from the "star average" of the conduction band occupations averaged over the 3 $\v k$-vectors related by the valley $C_3$ symmetry operations. A laser induced charge excitation exactly respecting the local valley symmetry, i.e. a charge excitation with exactly $C_3$ symmetry, will have $D(\v k)=0$ for all $\v k$. In such a state the macroscopic current will identically vanish. Finite values of $D(\v k)$, on the other hand, indicate a breaking of valley symmetry by the light pulse and the possibility of an emergent current.


\begin{figure*}[t!]
\begin{center}
\includegraphics[width=0.87\textwidth]{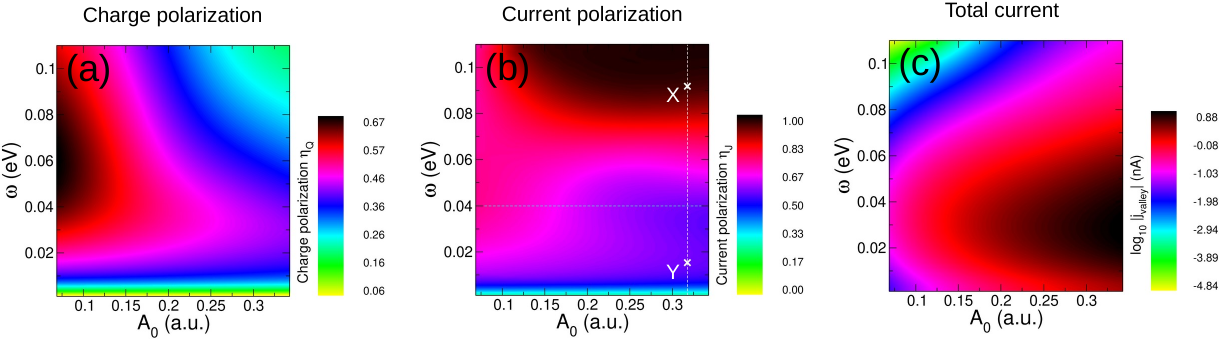}
\caption{\small {\it Maps of the charge and current excitation in the few cycle limit of circularly polarized light for graphene with a minimal 40~meV gap.} Fixing the duration of a THz pulse to X~fs we vary the amplitude, $A_0$, and frequency, $\omega$, evolving the lightform between single cycle and few cycle regimes. Shown are the valley polarization of the excited charge, panel (a), and of the emergent symmetry breaking current, panel (b), with the magnitude of this current presented in panel (c). High values of charge and current polarization are seen to occur at quite distinct pulse parameters, with gap tuned low amplitude light polarizing the charge response, but supra-gap and high amplitude light polarizing the current response. 
}
\label{fig3}
\end{center}
\end{figure*}

The vector potential of the light pulse tuned to a 30~meV reveals a pulse close to the single cycle regime, Fig.~\ref{fig2}(a), with the corresponding dynamical trajectory,  i.e. the evolution of crystal momenta this pulse induces, tracing out a single loop in momentum space, Fig.~\ref{fig2}(b). Failure of light-valley coupling can be seen in the momentum resolved charge excitations, Fig.~\ref{fig2}(c,d), which show charge excited at both valleys. The corresponding $D(\v k)$, however, also reveal a clear symmetry lowering of the excitation: the $C_3$ symmetry of the valley has been lowered to a $C_2$ symmetry in the light induced charge excitation, Fig.~\ref{fig2}(e,f). This arises as the momentum space loop shown in Fig.~\ref{fig2}(j) may evolve crystal momenta either towards the region of maximum coupling at the high symmetry K point, shown by the dipole matrix element plotted in (b), or away from it. Crystal momenta at positive $k_y$ (evolving towards K) will thus experience greater charge excitation than at negative $k_y$ (evolving away from K), leading to a charge imbalance at positive and negative $k_y$. It is this light-wave symmetry breaking, characteristic of the few cycle regime, that underpins the symmetry breaking of the charge density revealed by the $D(\v k)$ and the emergence of a non-vanishing current.

This appearance of current in the few cycle regime has been discussed recently in the context of large gap systems\cite{sharma_direct_2024}. However, while for large gap systems the currents generated are substantial and either fully spin and/or valley polarized, this is not the case here. The currents induced by light are both weak (in the pA regime) and, as inspection of the valley current polarizations shows, inset panel Fig.~\ref{fig1}(b), have significant valley polarization only in the weak tail of the current. It would thus appear that while the few cycle regime of excitation by THz circularly polarized light is endowed with an emergent current, the valley polarization of this is significantly degraded as compared that found in the visible spectrum gaps of the TMDC family.


To further explore this short time current response we lift the restriction on the light pulse that it must be gap tuned. The broadband nature of lightforms approaching the few optical cycle regime allows significant charge excitation for pulse central frequencies both significantly below and above the gap energy, permitting the sensible lifting of this restriction. In Fig.~\ref{fig3} we map the charge and current response of minimally gapped graphene as a function of the pulse frequency, $\omega$, and pulse amplitude, $A_0$. The band gap we take fixed at 40~meV, equal to that of a low misorientation angle graphene/h-BN system\cite{yankowitz_van_2019}, and fix the pulse duration to 117~fs, suitable for few cycle pulses in the THz window.

The valley polarization of excited charge reveals, at the low end of the range of pulse amplitudes explored, a pronounced maxima close to the band gap of 40~meV, Fig.~\ref{fig3}(a). While values of 67\% valley charge polarization can be achieved, this represents the valley signal at $T=0$; increasing the temperature rapidly induces significant charge excitation across the gap, washing out this valley contrast. Increase of $A_0$ results in significant weakening of this $T=0$ valley contrast, a feature that follows from the increase in $A_0$ generating dynamical trajectories that evolve crystal momenta further from the conduction band edge at which light-valley coupling holds, with degradation of valley polarization the inevitable consequence. 
From the current polarization, Fig.~\ref{fig3}(b), a dramatically different picture emerges. Once again a pronounced maxima in valley polarization exists, but now both shifted to energies significant above the gap and to high end of the range of pulse amplitudes explored. Strikingly, this maxima entails nearly 100\% valley polarized current with, moreover, current magnitudes now on the nA scale, Fig.~\ref{fig3}(c).

\begin{figure}[t!]
\begin{center}
\includegraphics[width=0.48\textwidth]{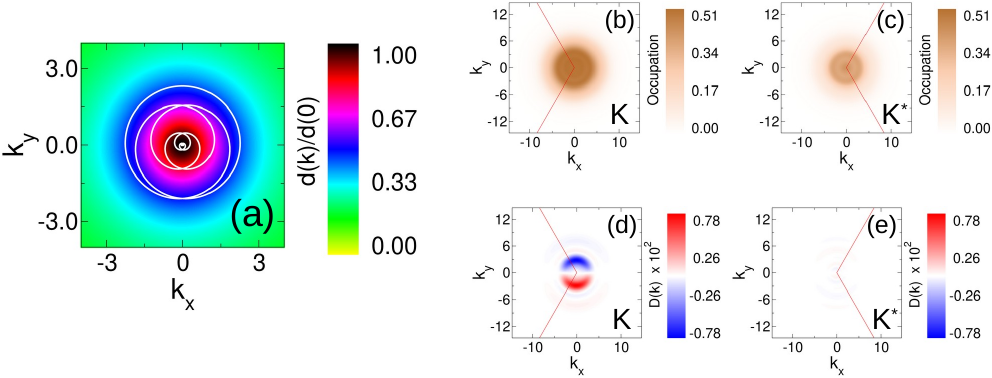}
\caption{\small {\it Valley selective symmetry breaking of charge excitation.} The dynamical trajectory of a few cycle circularly polarized light pulse, panel (a), consists of a series of loops that are successively dominant at positive or negative momenta relative to the initial $t=0$ crystal momenta. While charge excitation is seen at both K and K$^\ast$, the symmetries of the charge excitations are distinct: $C_2$ at the K valley, and $C_3$ at the K$^\ast$ valley, as revealed by the symmetry breaking densiry $D(\v k)$, Eq.~\ref{DK}, that takes on a finite dipole structure at K but is  numerically zero at K$^\ast$. This generates the 100\% polarized valley current revealed by the "light maps" presented in Fig.~\ref{fig3}.
}
\label{fig4}
\end{center}
\end{figure}

\begin{figure*}[t!]
\begin{center}
\includegraphics[width=0.8\textwidth]{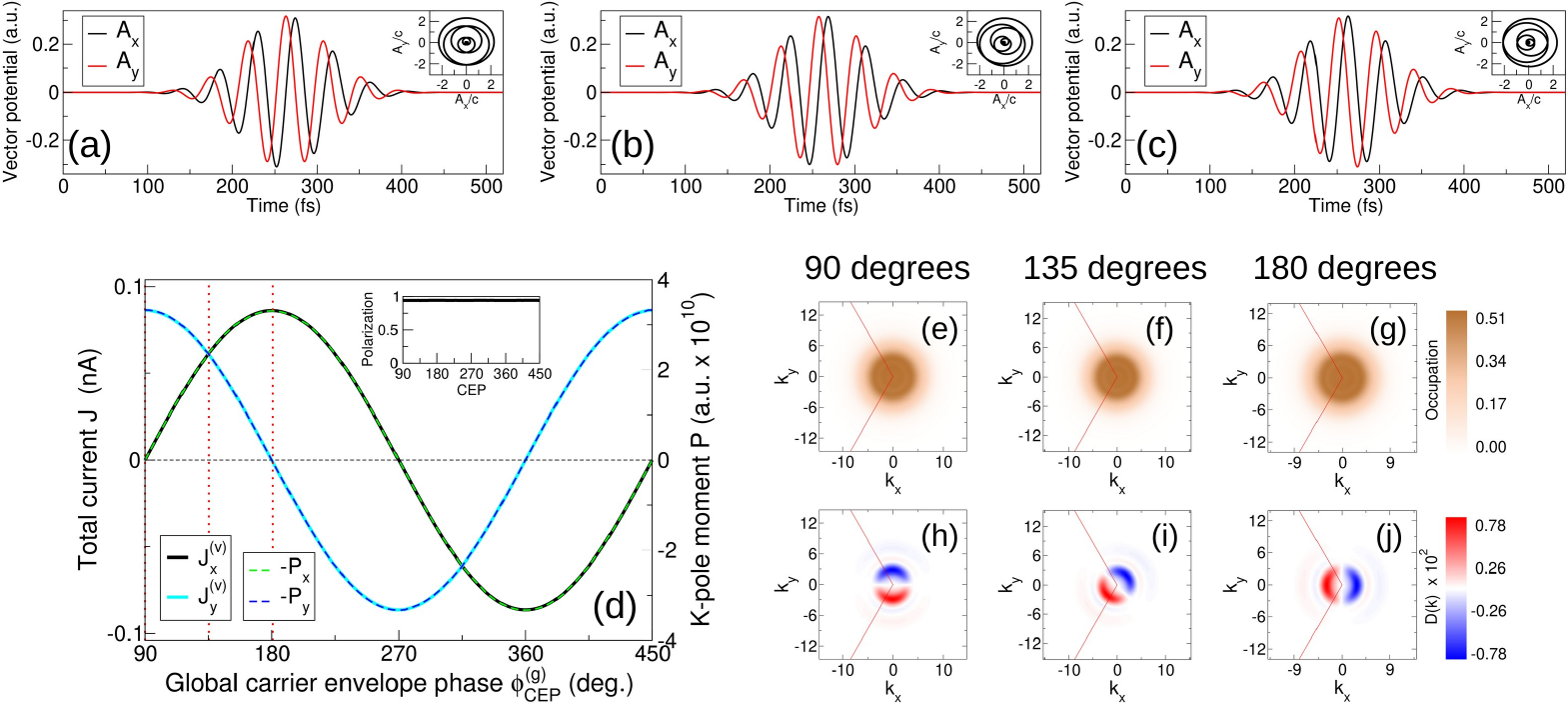}
\caption{\small {\it Valley current control via the global carrier envelope phase of few cycle circularly polarized light.} Few cycle circularly polarized light pulses are characterized by an orientable dynamical trajectory in momentum space, whose direction is determined by the global carrier envelope phase $\phi_g$, a phase shift common to both components of the light pulse, see Eq.~\ref{eqG}. Three representative examples are presented in (a-c), $\phi_g = 90^\circ$, $135^\circ$, and $180^\circ$; while the time evolving vector potentials are hard to distinguish the momentum space trajectories (inset panels) exhibit clear rotation with changing $\phi_g$. The direction of the valley polarized current follows exactly $\phi_g$, with the nearly 100\% value of valley polarization unchanged by variation of $\phi_g$. The charge excitation and symmetry breaking density for the three light pulses presented in (a-c) are shown in panels (e-j), as labelled, revealing the origin of current control via rotation of the symmetry breaking "K-pole" induced by few cycle circularly polarized light.
}
\label{fig5}
\end{center}
\end{figure*}

The supra-gap frequency at which this high polarization of current occurs implies a light pulse of several cycles. As may be seen in Fig.~\ref{fig4}(a) the dynamical trajectory, plotted for the pulse indicated by "X" in Fig.~\ref{fig3}(b), consists of several loops that, successively, evolve dominantly at either $k_y > 0$ or $k_y < 0$ momenta. Following the same arguments that applied to a single low symmetry loop, each of these loops therefore excites successively positive and negative contributions to the overall current induced by the light pulse, leading to the possibility of an overall cancellation. How this impacts the overall charge excitation can be seen in the momentum resolved excited charge and the corresponding symmetry breaking density $D(\v k)$, Fig.~\ref{fig4}(b,c) and (d,e) respectively, revealing that while charge is excited at both valleys only at the K valley is the symmetry breaking K-pole associated with the generation of valley current seen. The existence of a 100\% supra-gap valley polarized current response is thus seen to be a feature arising from cancellation over several loops of a few cycle pulse.


Essential to any viable scheme of lightwave valleytronics in graphene is the ability to control the laser pulse induced valley current. The "light maps" presented in Fig.~\ref{fig3} make clear that the magnitude of the valley current can be controlled both through the amplitude and frequency of the few cycle circularly polarized THz pulse. To see how the direction of the light generated current can be controlled we recall that a circular polarized light waveform can be expressed as

\begin{equation}
\v A(t) = f(t) (\sigma \cos(\omega t + \phi_g),\sin(\omega t + \phi_g))
\label{eqG}
\end{equation}
with $f(t)$ an envelope function encoding the pulse shape and temporal duration, and $\sigma=\pm 1$ the pulse helicity. At long pulse times the phase $\phi_g$ will play no physical role in light-matter interaction: only the carrier envelop phase difference of $\sigma \pi/2$ that determines the helicity, expressed as the sine and cosine functions in Eq.~\ref{eqG}, is of importance. However as the pulse waveform approaches the few cycle limit $\phi_g$ determines the direction of the orientable loop that the momentum space trajectory limits to. Three representative cases of this short time feature are shown in Fig.~\ref{fig5}(a-c), for $\phi_g = 90^\circ$, $135^\circ$, and $180^\circ$ respectively, where for vector potential and frequency we employ the pulse parameters of the point indicated by "X" in the light maps, Fig.~\ref{fig3}(a-c).

This provides an emergent short time pulse parameter that couples few cycle circularly polarized light to the direction of the induced current, and facilitates a complete direction control over current, Fig.~\ref{fig5}(d). Note that here we show the intraband component of the total current; for the pulses considered here the interband component is negligible in the residual post-pulse current, see Supplemental. The corresponding momentum resolved charge densities and $D(\v k)$ function (Eq.~\ref{DK}), panels (e-j), reveal that while the excited charge distribution does not discernibly change, this masks a clearly changing the K-pole that rotates with the global carrier envelope phase. Assigning a dipole moment to this K-pole leads, up to a scale, to exactly the same function as the valley current, as shown in Fig.~\ref{fig5}(d).


Opening a gap in graphene without significantly altering the Dirac cone electronic structure implies meV scale gaps, and hence light activation by THz pulses. While the light-valley coupling mechanism that underpins lightwave valleytronics in the transition metal dichalcogenide family holds also for minimally gapped graphene, the charge excitation induced by THz light is comparable to room temperature thermal excitation across the gap, washing out the valley signal and appearing to preclude a useful "THz valleytronics" in minimally gapped graphene. However, while valley charge cannot be effectively controlled circularly polarized THz light, it turns out that a second property of circularly polarized light, the short time emergence of a light-matter symmetry breaking current, does hold for gaps in the THz spectrum.

At supra-gap frequencies at which the excited charge exhibits only weak valley contrast, few cycle pulses of THz circularly polarized light generate a 100\% valley polarized current, with the THz polarization vector allowing full control over current direction by light. The emergence of a distinct region of light-matter coupling at which valley current control exists in the absence of valley charge polarization highlights a characteristic of few cycle circularly polarized light: it both possess a scalar degree of freedom -- the helicity that couples to valley charge just as for longer time pulses, but also a separate vectorial degree of freedom -- the orientation of the momentum space dynamical trajectory -- that couples to valley current. In the context of minimally gapped graphene this allows a rich current response in the few cycle regime of THz light, opening a route to the creation and control of valley currents and a "THz valleytronics" in this material.



\section*{Acknowledgements}

Sharma would like to thank DFG for funding through project-ID 328545488 TRR227 (projects A04), and Shallcross would like to thank DFG for funding through project-ID 522036409 SH 498/7-1. Sharma and Shallcross would like to thank the Leibniz Professorin Program (SAW P118/2021). The authors acknowledge the North-German Supercomputing Alliance (HLRN) for providing HPC resources that have contributed to the research results reported in this paper.



\section*{Additional information}

\noindent\textbf{Competing interests} The authors declare no competing financial, or any other, interests.

\noindent \textbf{Correspondence} Correspondence and requests for materials may be addressed to either of the corresponding authors.

\appendix

\section{Methods}

We model gapped graphene via the simplest nearest neighbour tight-binding model with a sub-lattice symmetry breaking gap $\Delta$

\begin{equation}
H = \begin{pmatrix}
\Delta/2 & f_{\v k} \\
f^\ast_{\v k} & -\Delta/2
\end{pmatrix}
\end{equation}
with $f_{\v k} = -t \sum_j e^{i\v k.\m\nu_j}$; here $t$ represents the nearest neighbour hopping, and $j$ is a sum over nearest neighbour vectors $\m\nu_j$.

In all calculations the laser pulse is described by a series of waveforms with Gaussian envelope form and with pulse centre $t_0^{(i)}$:

\begin{equation}
\v A(t) = \sum_i \v A_0^{(i)} \exp\left(-\frac{(t-t_0^{(i)})^2}{2\sigma_i^2}\right) \sin(\omega_i t + \phi_{cep}^{(i)})
\label{pulse}
\end{equation}
where $\v A_0^{(i)}$ is the pulse amplitude vector, $\sigma_i$ is related to the full width half maximum by $FWHM = 2\sqrt{2\ln{2}}\sigma_i$, $\omega_i$ is the frequency of the light, and $\phi_{cep}^{(i)}$ the carrier envelope phase.

The initial state is provided by a Fermi-Dirac distribution, and we include scattering that induces a loss of coherence in the simplest phenomenological model in which the density matrix is expressed in the eigenbasis at $\v k(t)$, with an exponential decay of the off-diagonal density matrix elements that encode quantum interference:

\begin{equation}
\partial_t \rho = -i\left[H,\rho\right] + \frac{1}{T_{D}} (\rho-\text{Diag}[\rho])
\label{LVN}
\end{equation}
where $\text{Diag}[\rho]$ denotes the matrix comprising only the diagonal elements of the density matrix $\rho$, and $T_{D}$ is a phenomenological decoherence time. For time propagation we employ the standard fourth-order Runge-Kutta method.

\bibliography{current,current1,samItems,salvini}

\end{document}